\documentclass[a4paper,11pt]{article}
\pdfoutput=1 

\usepackage{jcappub} 
\usepackage[utf8]{inputenc}
\usepackage{graphics}
\usepackage{epstopdf}
\usepackage{hyperref}
\usepackage{amsmath}
\usepackage{bigints}
\usepackage{relsize}
\usepackage{psfrag}
\usepackage{epsfig}
\usepackage{subfig}
\usepackage{xcolor}
\usepackage{url}

\usepackage[T1]{fontenc} 
\usepackage[percent]{overpic}
\usepackage{breqn}

 \normalsize
\newcommand{\bmat}{\left(\begin{array}}
\newcommand{\emat}{\end{array}\right)}
\newcommand{\be}{\begin{equation}}
\newcommand{\ee}{\end{equation}}
\newcommand{\bea}{\begin{eqnarray}}
\newcommand{\eea}{\end{eqnarray}}

\usepackage{booktabs}
\usepackage{siunitx}
\def\lsim{\raise0.3ex\hbox{$\;<$\kern-0.75em\raise-1.1ex\hbox{$\sim\;$}}}
\def\gsim{\raise0.3ex\hbox{$\;>$\kern-0.75em\raise-1.1ex\hbox{$\sim\;$}}}

\title{\boldmath Inflation, superheavy metastable strings and gravitational waves in non-supersymmetric flipped SU(5)}

\author[a]{George Lazarides,}
\author[b]{Rinku Maji,}
\author[c]{Ahmad Moursy,}
\author[d]{and Qaisar Shafi,}

\affiliation[a]{School of Electrical and Computer Engineering, Faculty of Engineering, \\
Aristotle University of Thessaloniki, Thessaloniki 54124, Greece}
\affiliation[b]{Laboratory for Symmetry and Structure of the Universe, Department of Physics,\\
		  Jeonbuk National University, Jeonju 54896, Republic of Korea}
\affiliation[c]{Department of Basic Sciences, Faculty of Computers and Artificial Intelligence,  \\
Cairo University, Giza 12613, Egypt}
\affiliation[d]{Bartol Research Institute, Department of Physics and Astronomy, \\
	University of Delaware, Newark, DE 19716, USA}
\emailAdd{glazarid@gen.auth.gr}
\emailAdd{rinkumaji9792@gmail.com}
\emailAdd{a.moursy@fci-cu.edu.eg}
\emailAdd{qshafi@udel.edu}
\abstract{Motivated by the NANOGrav 15 year data and other recent investigations of stochastic gravitational background radiation based on pulsar timing arrays, we show how superheavy strings survive inflation but the slightly heavier monopoles do not in a non-supersymmetric hybrid inflation model based on flipped $SU(5)$. With the dimensionless string tension parameter $G \mu\sim  10^{-6}$, the gravitational wave spectrum emitted by the strings, which are metastable due to breaking caused by monopole-antimonopole quantum mechanical tunneling, is compatible with the latest NANOGrav measurement as well as the advanced LIGO-VIRGO third run data. The string network undergoes about 30 $e$-foldings of inflation which suppresses the spectrum in the LIGO-VIRGO frequency range.
 With the symmetry breaking chain $SU(5) \times U(1)_X \to SU(3)_c \times SU(2)_L\times U(1)_Z \times U(1)_X \to SU(3)_c \times SU(2)_L \times U(1)_ Y$, the estimated proton lifetime is of order $10^{36}-10^{37}$ yrs.}
\begin{document}
\maketitle
\flushbottom
\section{Introduction}
\label{sec:intro}
The latest data released by NANOGrav \cite{NANOGrav:2023gor,NANOGrav:2023hvm} and other pulsar timing array experiments \cite{EPTA:2023fyk,Reardon:2023gzh,Xu:2023wog} have intensified exploration of the gravitational radiation emitted by topologically unstable superheavy cosmic strings. In Ref. \cite{Lazarides:2023ksx}, for instance, it is shown that the radiation emitted from quasi-stable superheavy strings with a dimensionless string tension $G\mu = 10^{-6}$ is consistent with the NANOGrav 15 year data. It is pointed out in Refs. \cite{Lazarides:2023ksx,Maji:2022jzu} that a late time inflationary phase is required in this case for consistency with the third run advanced LIGO-VIRGO data \cite{LIGOScientific:2021nrg} in the higher frequency range. Walls bounded by superheavy strings provide another scenario \cite{Maji:2023fba,Lazarides:2023ksx} in good agreement with the NANOGrav data.
An alternative scenario is based on metastable strings with the dimensionless string tension parameter $G\mu \approx 10^{-7}$ \cite{NANOGrav:2023gor,Buchmuller:2023aus,Antusch:2023zjk,Fu:2023mdu}.\footnote{According to our own understanding the $SO(10)$ symmetry breaking chain described in Ref.~\cite{Fu:2023mdu} does not produce metastable strings.}
 The latter parameter is carefully chosen to explain the NANOGrav observations, without running into (serious) conflict with the LIGO-VIRGO results.

Quasi-stable and metastable strings have an important feature in common. They are both topologically unstable and arise from the symmetry breaking ${\cal G} \to {\cal H} \times U(1)$, which produces monopoles, and the subsequent $U(1)$ breaking yielding strings (flux tubes) which connect monopoles with antimonopoles. In the quasi-stable case, the monopoles and antimonopoles undergo partial inflation, connect to strings, and re-enter the horizon. In the metastable case the monopoles and antimonopoles are inflated away, but can appear on strings via quantum pair creation. This last feature requires that, in the metastable case, the symmetry breaking scales associated with the monopoles and strings are quite close to each other.

In this paper we present a hybrid inflationary scenario based on a non-supersymmetric flipped $SU(5)$ model which successfully implements the metastable string scenario. The prediction for the scalar spectral index $n_s$ is in good agreement with the recent measurements \cite{Planck:2018jri,BICEP:2021xfz}, and the tensor-to-scalar ratio lies in the 
range $r\sim 10^{-4}-10^{-3}$. The observed baryon asymmetry can be realized either via thermal \cite{Fukugita:1986hr} or non-thermal leptogenesis \cite{Lazarides:1990huy}, but we will not pursue this in detail here. This metastable string model with $G\mu \sim 10^{-6}$ will be tested by the ongoing LIGO-VIRGO experiment. The estimated proton lifetime lies in the range $10^{36} - 10^{37}$ yrs.

This paper is organized as follows. In Sec.~\ref{sec:model2}, we provide the salient features of our flipped $SU(5)$ model, and in Sec.~\ref{sec:infdynamic} we summarize the symmetry breaking and outline the inflationary scenario. In Sec.~\ref{sec:observables}, we discuss the inflationary dynamics, observables and reheating. We investigate the Coleman-Weinberg 1-loop radiative correction to the tree level inflation potential in Sec.~\ref{sec:RCs} and show that, although in certain cases it is very small, it slightly improves their inflation observables.
Sec.~\ref{sec:string-GW}  provides details about the gravitational wave spectrum generated by the superheavy metastable strings that survive inflation. In particular, we are able to show how $G\mu\sim 10^{-6}$ can be reconciled with the third advanced LIGO-VIRGO run \cite{LIGOScientific:2021nrg}. This comes about because the string network in this case undergoes around 30 $e$-foldings of inflation. In Sec.~\ref{sec:proton_life_time}, we estimate the proton lifetime and show that it is well above the bound from the Super-Kamiokande experiment \cite{Super-Kamiokande:2020wjk}.
Finally, our conclusions are summarized in Sec.~\ref{sec:conc}.
%
\section{Inflation and metastable strings in flipped $SU(5)$}
\label{sec:model2}
We provide an inflation model with the gauge symmetry being $SU(5)\times U(1)_X$ (flipped $SU(5)$), one of the maximal subgroups of the $SO(10)$ gauge group. In order to realize metastable strings in flipped $SU(5)$, we consider the following  symmetry breaking chain (for a recent discussion of topological structures in SO(10) and its various subgroups see Ref.~\cite{Lazarides:2023iim}, whose notation we follow here):
\bea \label{eq:SBpatternX} 
SU(5)\times U(1)_X  &\xrightarrow[]{\left<\Phi\right>} &  SU(3)_c \times SU(2)_L \times U(1)_{Z}  \times U(1)_X \nonumber \\
&\xrightarrow[]{\left<\Psi\right>} &  SU(3)_c \times SU(2)_L \times U(1)_{Y} .
\eea
%
\begin{table}[h!]
 \centering
 \begin{tabular}{c | c c c c c c c }
  \hline \hline
   &$f$ & $F$ & $e^c$ & $h$ & $\Phi$ & $\Psi$  & $S$ \\
   \hline 
     $ SU(5) $  & $\mathbf{\bar 5}_{F}$ & $\mathbf{10}_{F}$  & $\mathbf{1}_F $  & $\mathbf{\bar 5}_{H}$ & $\mathbf{24}_{H}$ & $\mathbf{10}_{H}$ & $\mathbf{1}_H $  \\
   \hline 
   $U(1)_{X}$ & 3  & -1 & -5 & -2 & 0 & -1  & 0 \\
   \hline  \hline
  \end{tabular}
 \caption{{Representations and $U(1)_X$ charge assignments of the matter and Higgs fields in $SU(5)\times U(1)_X$.}}
 \label{tab:FSU5}
\end{table}

The representations used in the minimal flipped $SU(5)$ are shown in Table~\ref{tab:FSU5} \cite{Kyae:2005nv,Moursy:2020sit,Lazarides:2023iim}:
\begin{itemize}
\item The fifteen chiral fermions of the Standard Model (SM) plus the right-handed neutrino, per family, are accommodated in the following representations: $F=\mathbf{10}_{F}(-1)= \{ Q, d^c, \nu^c \}$, $f=\mathbf{\overline{5}}_{F}(3)=\begin{pmatrix}
u^c \\
L
\end{pmatrix} $ and $\mathbf{1}_{F}(-5)=e^c$.

\item The electroweak Higgs doublet  is contained in $h=\mathbf{\overline{5}}_{H}(-2)
=\begin{pmatrix}
D^c \\
H
\end{pmatrix}. 
$
\item The adjoint representation $\Phi=\mathbf{24}_{H}$ triggers the first stage of symmetry breaking, $SU(5) \to SU(3)_c \times SU(2)_L \times U(1)_Z\times U(1)_X $, by acquiring a vacuum expectation value (vev) along the $G_{\rm SM}\times U(1)_Z$ neutral direction.

\item The representation $\Psi= \mathbf{10}_{H}(-1)= \{ Q_H, d_H^c, \nu_H^c \} $ triggers the second breaking stage, $U(1)_Z\times U(1)_X \to U(1)_Y$, by acquiring a vev in the SM neutral direction ${\nu}^c_H$.
\item The  inflaton corresponds to the real singlet $S=\mathbf{1}_{H}$. 
\end{itemize}

The first breaking yields magnetic monopoles carrying $SU(3)_c$, $SU(2)_L$ and $U(1)_Z$ magnetic fluxes. The breaking of $U(1)_ Z \times U(1)_X$ in the next step to $U(1)_Y$ produces strings that can connect the monopoles and antimonopoles. The metastable string scenario for producing gravitational waves compatible with the NANOGrav 15 year data and the LIGO-VIRGO third run results requires that the strings are superheavy with the dimensionless string tension parameter $G\mu$ of order $10^{-6}$. The associated monopoles produced in the first breaking  should then be only slightly heavier. Furthermore, it is required that the primordial monopoles are inflated away but certainly not the strings that provide the gravitational waves. The metastability of the strings arises from the quantum mechanical production of monopole-antimonopole pairs.

Our task is to show how this can be realized in a hybrid inflationary scenario based on the non-supersymmetric flipped $SU(5)$ model with $G\mu \sim  10^{-6}$.
We will inflate using the singlet real scalar field $S$. The gauge invariant terms in the scalar potential relevant for inflation are as follows:
\bea\label{eq:pot_tot}
V &\supset & V_0 - \mu_\Phi^2 \, tr(\Phi^2) -  \dfrac{\mu_1}{3}  \, tr(\Phi^3) + \dfrac{\lambda_1}{4} \, tr(\Phi^4) + \lambda_2 \, [tr(\Phi^2)]^2
- \dfrac{\mu_\Psi^2}{2} \, tr(\Psi^\dagger\Psi) + \dfrac{\lambda_3}{4} \, [tr(\Psi^\dagger\Psi)]^2  \nonumber \\ 
&& + \dfrac{ \lambda_4}{4} \, tr(\Psi^\dagger\Psi\Psi^\dagger\Psi) + \lambda_5 \, tr(\Psi^\dagger \Phi^2 \Psi)+ \lambda_6 \, tr(\Psi^\dagger  \Psi) \, tr(\Phi^2) + \mu_2 \, tr(\Psi^\dagger\Phi\Psi)\nonumber\\
&& +\dfrac{m^2}{2} S^2 
+  \lambda_7 S^2 tr(\Psi^\dagger  \Psi)  -  \lambda_8 S^2 tr(\Phi^2)    \, ,  
\eea
where the adjoint representation $\Phi^\alpha_{\,\, \beta}\equiv \phi_a(T^a)^\alpha_{\, \beta}$, with $T^a$ being the $SU(5)$ generators, and the 10-plet $\Psi$ is a 
$5\times 5$ complex antisymmetric matrix $\Psi^{\alpha\beta}$. Here, we use the indices $a,b,c, \cdots=1,2,\cdots 24$, and $\alpha, \beta, \cdots=1,2,\cdots 5$, and the sum over repeated indices is understood implicitly. We have not provided here the terms involving the 5-plet higgs which is frozen at the origin during inflation. However, it plays a role during reheating as we shall see later. 
To implement hybrid inflation, we do not consider a variety of other terms such as the linear, cubic and quartic terms in $S$  as well as many others, assuming they can be safely ignored.
The constant vacuum energy $V_0$ is added such that the potential is zero at the true minimum.

For suitably large $S$ field values, the 10-plet Higgs field $\Psi$ plays the role of the waterfall field that is frozen at the origin during inflation until $S$ reaches a critical value $S_c$ at which the waterfall phase transition of the hybrid inflation scenario is triggered. On the other hand, the  24-plet Higgs field $\Phi$ follows a field dependent minimum during inflation, and finally falls into its true minimum at a scale $\sim 10^{16}$ GeV, which is the scale associated with the monopoles.
Thus, $SU(5)$ is broken during and after inflation, and hence the monopoles are inflated away. Denoting $ \phi \equiv \phi_{24}$, and the vev $ \langle \phi \rangle={v_\phi} $, we have
\be 
\langle \Phi \rangle=  \dfrac{v_\phi}{\sqrt{15}} {\rm diag}(1,1,1,-3/2,-3/2).
\ee
The breaking of $U(1)_Z\times U(1)_X$ to $U(1)_Y$ occurs after $S$ reaches $S_c$, caused by the 10-plet higgs vev in the SM neutral direction $\nu_H^c$, at a scale close to the $SU(5)$ breaking scale. The metastable cosmic strings may experience a large number of $e$-foldings, and eventually produce a stochastic gravitational wave spectrum that is consistent with the 15 year NANOGrav data and the third advanced LIGO-VIRGO bound. Rotating $\Psi^{45} $ to the real axis by a $U(1)_Z$ transformation, we 
define the normalized real scalar field $\psi \equiv \nu^c$ whose vev $\langle \psi\rangle=v_\psi$ 
is the cosmic string scale.

Clearly, the couplings of the inflaton field $S$ to $\Psi$ and $\Phi$ play a crucial role in realizing the above scenario. Indeed, the  $\lambda_7$ and $\lambda_8$ terms significantly modify the standard hybrid inflation tree level potential \cite{Linde:1993cn}, and enable us to have a hill-top shape potential, as advocated in Ref. \cite{Ibrahim:2022cqs}, that we discuss in the next section.
%
\section{Symmetry breaking and inflation}
\label{sec:infdynamic}
For simplicity, we set the coefficients $\mu_1=\mu_2=0$ in Eq.~(\ref{eq:pot_tot}), and assume that all remaining coefficients are real. The inflationary potential then takes the form
\bea\label{eq:pot_inf}
V_{\rm inf} &=& V_0 - \dfrac{m_\phi^2}{2} \, \phi^2  + \dfrac{\beta_\phi}{4} \, \phi^4
- \dfrac{m_\psi^2}{2} \, \psi^2 + \dfrac{\beta_\psi}{4} \, \psi^4  + \frac{\beta_{\psi\phi}}{2} \, \psi^2\phi^2 +\dfrac{m^2}{2} S^2 
+  \frac{\beta_{S\psi}}{2} \, S^2 \, \psi^2  -  \frac{\beta_{S\phi}}{2} \,  S^2 \, \phi^2   \, , \nonumber \\
\eea
 with the remaining components of $\Phi$ and $\Psi$ fixed at zero during and after inflation. The parameters $m_\phi, m_\psi, \beta_\phi, \beta_\psi,\beta_{\psi\phi} , \beta_{S\phi}, \beta_{S\psi}$ are given in terms of the parameters of the original potential in Eq.~(\ref{eq:pot_tot}) as follows:
\bea
\beta_\phi &=& \frac{1}{120} (7 \lambda_1+120 \lambda_2),  \hspace{1cm} \hspace{1cm} \beta_{\psi}={\lambda_3}+\frac{\lambda_4}{2} , \hspace{1cm} \beta_{\psi\phi}= \frac{3\lambda_5}{10} + \lambda_6  \, , \nonumber\\
\beta_{S\psi} &=& 2 \lambda_7 
 , \hspace{1cm} \beta_{S\phi} = \lambda_8 , \hspace{1cm} m_\phi = \mu_\phi ,   \hspace{1cm}  m_\psi=\mu_\psi.
\eea
 The vevs of the scalar fields $S$,$\phi$ and $\psi$ at the true minimum of the potential are respectively given by
 \bea
 v_S=0, \hspace{1cm}  v_\phi= \sqrt{\frac{\beta _{\psi } m_{\phi }^2-{m_\psi}^2 \beta _{\psi \phi }}{\beta _{\psi } \beta _{\phi }-\beta _{\psi \phi }^2}} , \hspace{1cm}  v_\psi= \sqrt{\frac{\beta_\phi  m_{\psi }^2-\beta _{\psi \phi } m_{\phi }^2}{\beta _{\psi } \beta _{\phi }-\beta _{\psi \phi }^2}}.
 \eea
 The mass squared of the scalar field $S$ at the true minimum is
 \bea
M_S^2= m^2+\frac{m_{\psi }^2 \left(\beta _{\phi } \beta _{\text{s$\psi $}}+\beta _{\psi \phi } \beta _{\text{s$\phi $}}\right)-m_{\phi }^2 \left(\beta _{\psi \phi } \beta _{\text{s$\psi $}}+\beta _{\psi } \beta _{\text{s$\phi $}}\right)}{\beta _{\phi } \beta _{\psi }-\beta _{\psi \phi }^2},
\eea
while the mass-squared matrix at the true minimum, in the basis of $(\psi,\phi)$, is given by
\bea
{\cal M}^2= 
\left(
\begin{array}{cc}
 2 \beta _{\psi } v_\psi^2 & 2 \beta _{\psi \phi } v_\psi v_\phi  \\
2 \beta _{\psi \phi } v_\psi v_\phi  & 2 \beta _{\phi }  v_\phi^2  \\
\end{array}
\right).
\eea
%
%
 
 Minimizing the potential in Eq.~(\ref{eq:pot_inf}) in the $\psi$ and $\phi$ directions, we find that $\psi$ is fixed at zero during inflation until $S$ reaches $S_c$, while $\phi $ has an inflaton dependent minimum. Therefore, the trajectory in the $(\psi,\phi)$ plane is given as follows:\footnote{For related discussions in supersymmetric models, see for instance Refs. \cite{Jeannerot:2000sv,Lazarides:2008nx}.}
\bea\label{eq:trajec} 
(\psi,\phi) = \left(0 \, ,\,  \sqrt{\frac{m_{\phi }^2+  \beta _{S \phi } S^2 }{\beta _{\phi }}} \right) \,.
 \eea
Thus, with non-zero values of $\phi$ during inflation, the monopoles are inflated away.
The field dependent squared-mass matrix in the basis $(S,\phi,\psi)$  during inflation is given by
\bea\label{eq:infmass2}
\!\!\!\!\!\!\!\!\!M^2_{\rm inf}= 
\left(
\begin{array}{ccc}
 m^2-\frac{ \beta_{S\phi}  \left(m_{\phi }^2+ \beta _{S\phi} \, S^2\right)}{\beta _{\phi }} & -\frac{2 S \, \beta _{S\phi} \sqrt{m_{\phi }^2+ \beta _{S\phi}\, S^2 }}{\sqrt{\beta _{\phi }}} & 0 \\
 -\frac{2 S \,\beta _{S\phi} \sqrt{m_{\phi }^2+ \beta _{S\phi}\, S^2 }}{\sqrt{\beta _{\phi }}} & 2 \left(m_{\phi }^2+ \beta _{S\phi}\,S^2 \right) & 0 \\
 0 & 0 & -m_{\psi }^2+ \beta _{S\psi}S^2 +\frac{\beta _{\psi \phi } \left(m_{\phi }^2+ \beta _{S\phi} \, S^2\right)}{\beta _{\phi }}\\
\end{array}
\right). 
\eea
The  mass-squared element $(M^2_{\rm inf})_{\psi\psi}$ flips its sign when the inflaton $S$ passes a critical value $S_c$, which is given by setting $(M^2_{\rm inf})_{\psi\psi}=0$. Thus,
\be 
S_c= \sqrt{\frac{\beta _{\phi } m_{\psi }^2-\beta _{\psi \phi } m_{\phi }^2}{\beta _{\psi \phi } \beta _{{S\phi }}+\beta _{\phi } \beta _{{S\psi }}}} \,.
\ee
For $S > S_c$, $(M^2_{\rm inf})_{\psi\psi} >0$, and hence $\psi$ stays at  the origin, but for $S< S_c$, $(M^2_{\rm inf})_{\psi\psi}<0$, and hence the  waterfall phase is triggered at $S=S_c$ with the generation of cosmic strings. 

The parameter space in our scenario allows for both a prompt waterfall~\cite{Linde:1993cn}, for which inflation ends at $S=S_c$, and a mild waterfall with inflation ending when the fields start to oscillate after undergoing a relatively large number of $e$-foldings between the start of waterfall and the onset of scalar fields oscillations~\cite{Clesse:2010iz, Kodama:2011vs, Clesse:2015wea}. In this paper we focus on an intermediate case such that the cosmic strings formed at the cosmic time when $S=S_c$ are partially inflated by fewer $e$-foldings between the start of waterfall and the end of inflation. 
In the valley in Eq.~(\ref{eq:trajec}), a single field inflation can be realized with a tree level effective potential of the form \cite{Ibrahim:2022cqs,Rehman:2009wv}
\bea\label{eq:infpot1}
 V_{\text{inf}}(\widetilde{S})= \widetilde{V}_0\left( 1+  \widetilde{S}^2- \gamma \, \widetilde{S}^4\right),
\eea
where we have used the following redefinitions \cite{Ibrahim:2022cqs,Rehman:2009wv}
\bea 
 \widetilde{V}_0 \equiv V_0-\dfrac{m_\phi^4}{4\beta_\phi} 
 \,, \hspace{0.5cm} \widetilde{S} \equiv \sqrt{\dfrac{\eta_0}{2}} \, S
 \,, \hspace{0.5cm} \eta_0 \equiv \frac{m^2 \beta _{\phi }-m_{\phi }^2 \beta _{S\phi }}{\widetilde{V}_0 \,\beta _{\phi }} 
 \,, \hspace{0.5cm} \gamma \equiv \frac{\beta _{S \phi }^2}{\eta _0^2 \widetilde{V}_0 \,\beta _{\phi }} \,.
\eea 
The Hubble parameter during inflation is then given by $H\approx \sqrt{V_{\text{inf}}/3 M_{\rm Pl}^2}$, where $M_{\rm Pl}$ is the reduced Planck scale.
In order to guarantee the stability of the inflation trajectories during inflation, we calculate $M^2_{\rm inf}$ after solving the equations of motion of the complete system, and make sure that the field dependent mass squared of $\psi$ and $\phi$ during inflation are much larger than $H^2$. Moreover, the mixing between $S$ and $\phi$ in the squared-mass matrix in Eq~(\ref{eq:infmass2}) is always very small and will not affect the slow rolling of $S$.
%
\section{ Inflationary dynamics, observables and reheating }
\label{sec:observables}
As advocated above, we focus on an intermediate case between prompt and very mild waterfall, where inflation continues for a relatively small number of $e$-foldings after crossing the instability point at $S_c$. In this case, the observable scales leave the Hubble radius when the fields are still evolving along the trajectory Eq.~(\ref{eq:trajec}), which allows us to  calculate the inflation observables using the single field slow-roll formalism. 

The inflation potential in Eq~(\ref{eq:infpot1}) has the hilltop shape with a local maximum (hilltop) located at 
\be
\widetilde{S}_{\rm m}=\pm\dfrac{1}{\sqrt{2\gamma}}.
\ee  
In order to obtain successful inflation, we should fulfill the condition $\widetilde{S}_c < \widetilde{S}_*< \widetilde{S}_{\rm m}$, where $\widetilde{S}_*$ is the value of $\widetilde{S}$ when the pivot scale $k_*$ exits the inflationary horizon. The inflaton will roll down from a value very close to $\widetilde{S}_{\rm m}$, and we have two cases, $\widetilde{S}_* \gg 1 $ (corresponding to $\gamma\ll 1)$, and $\widetilde{S}_* \ll 1$ (corresponding to $\gamma\gg 1)$. The slow-roll parameters are then given by
\bea
\epsilon=\frac{\eta_0}{4}\left({\frac{V_{\widetilde{S}}}{V}}\right)^2\,, 
\hspace{0.5cm}
\eta=\frac{\eta_0}{2}\left({\frac{V_{\widetilde{S}\widetilde{S}}}{V}}\right),
\eea
where we have dropped the subscript ``inf'' for simplicity, and we work in the units where $M_{\rm Pl}=1$. The total number of e-foldings $\Delta N_*$ between the time when the pivot scale $k_*=0.05 \,{\rm Mpc^{-1}}$ exits the horizon and the end of inflation is calculated from 
\bea\label{eq:Ns1}
\Delta N_{*}=\sqrt{\frac{2}{\eta_0}}\bigintss_{\widetilde{S}_{e}}^{\widetilde{S}_*}{\frac{d\widetilde{S}}{\sqrt{\epsilon(\widetilde{S})}}} ,
\eea
where  $\widetilde{S}_e$ is the $\widetilde{S}$ value at the end of inflation.
However, the number of e-foldings required to solve the horizon and flatness problems is alternatively calculated from the thermal history of the Universe \cite{Liddle:2003as,Chakrabortty:2020otp,Kawai:2023dac}: 
\begin{equation}\label{eq:Ns2}
\Delta N_* \simeq 61.5 + \frac{1}{2} \mathrm{ln} \frac{\rho_*}{M_{\rm Pl}^4}-\frac{1}{3(1+\omega_r)} \mathrm{ln} \frac{\rho_e}{M_{\rm Pl}^4} + \left(\frac{1}{3(1+\omega_r)} - \frac{1}{4} \right)\mathrm{ln} \frac{\rho_r}{M_{\rm Pl}^4} \ .
\end{equation}
Here $\rho_* = V(\widetilde{S}_*)$  is the energy density of the Universe when the pivot scale exits the horizon, $\rho_e = V(\widetilde{S}_e)$ is the energy density at the end of inflation, $\rho_r = (\pi^2/30) g_*T_r^4$ is the energy density at the reheating time, and $w_r$ is the effective equation-of-state parameter from the end of inflation until reheating that we set equal to zero~\cite{Senoguz:2015lba,Chakrabortty:2020otp}. An upper bound on the reheating temperature, $T_r\lesssim 10^{12}$ GeV is discussed below. 
 The effective number of massless degrees of freedom at the reheating time is taken  $g_*= 106.75$, corresponding to the SM spectrum. In our numerical analysis, we calculate $\Delta N_*$ such that its values from Eqs.~\eqref{eq:Ns1} and \eqref{eq:Ns2} coincide. 

We will denote the number of e-foldings between the times corresponding to $S_c$ and $S_e$ by $\Delta N_c$. We define 
$t_c$ to be the time when $S=S_c$, $t_e$ is the time at the end of inflation, and $t_r$ is the time at reheating, which is calculated from the relation \cite{Lazarides:1997xr,Lazarides:2001zd}
\be
T_r^2= \sqrt{\dfrac{45}{2\pi^2 \, g_*}} \, \dfrac{M_{\rm Pl}}{t_r}.
\ee 

The Cosmic Microwave Background (CMB)  observables including the scalar spectral index $n_s$, tensor-to-scalar ratio $r$ and the amplitude of scalar perturbations  $A_s$ are calculated at the horizon exit of the pivot scale as follows:
\bea 
n_s=1- 6 \epsilon_* + 2 \eta_* \,, \hspace{0.5cm}  r=16 \epsilon_* \,,  \hspace{0.5cm}   A_s=\frac{V_*}{24 \pi^2 \epsilon_*} .
\eea
\begin{table}[h!]
 \centering
 \begin{tabular}{c |  c c c  }
  \hline \hline
   & $\widetilde{V}_0[M_{\rm Pl}^4]$ & $\eta_0[M_{\rm Pl}^{-2}]$ & $\gamma$    \\
   \hline 
   BP1 &  $2.14 \times 10^{-11}$ & $0.013$ & $1.2$   \\
   \hline 
    BP2 &  $2.65 \times 10^{-11}$ & $0.013$ & $1.2$  \\
  \hline  \hline
  \end{tabular}
 \caption{{ Parameter values for the inflation potential in Eq.~(\ref{eq:infpot1}) for the two benchmark points BP1 and BP2 considered.}}
 \label{tab:par1}
\end{table}
\begin{table}[h!]
 \centering
\begin{tabular}{c |  c c c c c c c c}
  \hline \hline
   & $m$ & $m_\psi$ & $m_\phi$  & $\beta_\psi$ & $\beta_\phi$  & $\beta_{\psi\phi}$ & $\beta_{S\psi}$ &  $\beta_{S\phi}$\\
   \hline 
   {BP1} &  $1.45\times 10^{12}$ & $ 4\times 10^{15}$ & $1.94\times 10^{15}$  & 3.5 & 0.265 & $-0.92$  & $2.15\times 10^{-7}$  & $3.4\times 10^{-8}$ \\
  \hline  
  {BP2} &  $1.68 \times 10^{12}$ & $1.12\times 10^{15}$ & $1.94\times 10^{15}$  & $1.4$ & $0.118$  & $-0.4$ & $1.3 \times 10^{-7}$ &  $2.5\times 10^{-8}$ \\
   \hline  \hline
  \end{tabular}
 \caption{The parameters in Eq.~(\ref{eq:pot_inf}) for the benchmark points BP1 and BP2 yielding acceptable solutions for realizing metastable strings in hybrid inflation. Dimensionful parameters are given in GeV.}
 \label{tab:par2}
\end{table}
\begin{table}[h!]
 \centering
 \begin{tabular}{c |  c c c c c }
  \hline \hline
   & $v_\phi$ & $v_\psi$ & $M_\phi$  & $M_\psi$ & $M_S$  \\
   \hline 
   Obs(BP1) &  $1.9\times 10^{16}$ & $ 1\times 10^{16}$ & $3.5\times 10^{15}$  & $2.96 \times 10^{16}$ & $ 3.36 \times 10^{12}$  \\
   \hline 
   Obs(BP2) &  $ 3 \times 10^{16}$ & $ 1.6\times 10^{16}$ & $2.5 \times 10^{15}$  & $3 \times 10^{16}$ & $3.72\times 10^{12}$  \\
  \hline  \hline
  \end{tabular}
 \caption{ The physical masses and vevs are given in GeV, and BP1 and BP2 correspond to metastable strings with $G\mu =2.1 \times 10^{-6}$ and $G\mu = 5.4\times  10^{-6}$, respectively.}
 \label{tab:ob1}
\end{table}
\begin{table}[h!]
 \centering
 \begin{tabular}{c |  c c c c c c }
  \hline \hline
   & $A_s$ &  $n_s$ & $r$   & $\Delta N_*$ & $\Delta N_c$ & $S_*[M_{\rm Pl}]$ \\
   \hline 
   Obs(BP1) &  $2.18\times 10^{-9}$ & $ 0.963$ & $8\times 10^{-4}$  & $51.1$ &  $31$  & 7.405  \\
   \hline 
   Obs(BP2) &  $2.1\times 10^{-9}$ & $ 0.962$ & $1.0\times 10^{-3}$  & $53.4$ &  $31$  & 7.454  \\
  \hline  \hline
  \end{tabular}
 \caption{Predictions for CMB observables and number of $e$-foldings. $\Delta N_c$ denotes the number of e-foldings from the start of the waterfall transition until the end of inflation.}
 \label{tab:ob2}
\end{table}
%
%
%

The value of $\widetilde{V}_0$ is fixed by the observed value of $A_s= (2.099\pm0.101) \times 10^{-9}$~\cite{BICEP:2021xfz,Planck:2018jri}. We have explored both trans-Planckian and sub-Planckian values of the inflaton $S$
at the time of horizon exit of the pivot scale. As pointed out above, we have focused on the case where the pivot scale exits the horizon before the time when $S$ reaches $S_c$.
For $\widetilde{S}_*\gg 1$, we have a mild waterfall regime of hybrid inflation, where a large number of $e$-foldings greater than $60$ is found after the waterfall,\footnote{See Refs.~\cite{Clesse:2010iz,Kodama:2011vs,Clesse:2015wea} and the references therein, for more details.} which is not good for the inflationary observables ~\cite{Clesse:2010iz, Kodama:2011vs}. When $\tilde{S}_* \lesssim 1$, we may have both intermediate and prompt waterfall, and information about the parameters for two benchmark points suitable for our purposes are given in Tables~\ref{tab:par1} and~\ref{tab:par2}.\footnote{The detailed study of the whole parameter space is beyond the scope 
of this paper. However the most signifficant constraints for 
successful solutions arise from the inflation observables, as well as 
the requirement that the vevs of the monopole and waterfall fields $\phi$ and $\psi$ are of order $10^{16}$ GeV. Our scenario will of course have to confront future 
measurements.} 
The scalar field masses $M_\phi,M_\psi,M_S$ and higgs vevs, as well as the inflation observables, are given in Tables~\ref{tab:ob1} and \ref{tab:ob2}. Both the given benchmarks can explain the NANOGrav results. The inflation observables were calculated using the total inflation potential which includes the Coleman-Weinberg radiative correction that will be discussed in some detail in the next section. The values of the observables calculated after adding the latter correction are very close to the ones computed from the tree level inflation potential in Eq.~(\ref{eq:infpot1}),  which they only slightly improve.

Let us briefly discuss how reheating occurs in this hybrid inflation model from the oscillations of the inflaton field $S$ and the “monopole” and “waterfall” fields $\phi$ and $\psi$ around their respective minima. 
The gauge singlet scalar $S$ and the adjoint scalar $\phi$ have effective trilinear couplings to the SM Higgs doublet. Since the discussion is similar for the two fields, we focus on the $S$ field with the trilinear coupling given by $\delta \, H^\dagger   H \,  S$, where the dimensionful parameter $\delta \lesssim M_S$, the inflaton mass, in order to retain perturbativity. This coupling yields a reheat temperature $T_r \lesssim 10^{13} \times ( \delta / M_S)$ GeV, with $M_S \sim 10^{12}$ GeV. Thus, for $( \delta / M_S) \lesssim 1/10$, the reheat temperature associated with this decay is less than or of order $10^{12}$ GeV. A similar discussion holds for the decay of the adjoint scalar field $\phi$.

Regarding the waterfall field $\psi$, consider the non-renormalizable coupling
$\frac{f}{ M_{\rm Pl}} 10_F 10_F 10_H^\dagger \\ 10_H^\dagger$, which provides masses to the right handed neutrinos, where $f$ denotes a dimensionless coupling and, for simplicity, we have left out the family indices. This interaction yields an effective renormalizable Yukawa coupling of the waterfall field to the right handed neutrinos which is given by  $(M_R / v_\psi) 10_F \, 10_F \, \psi$, where $M_R$ denotes the relevant right handed neutrino mass. From this interaction the reheat temperature is estimated to be
of order $10^{16} \times ( M_R / v_\psi)$ GeV. 
With the term in the last bracket $\lesssim 10^{-4}$ as the right-handed neutrino masses are typically of intermediate scale, we again find a 
reheat temperature $\lesssim 10^{12}$ GeV.

The benchmark point BP1 corresponds to trans-Planckian values of the inflaton $S$, but interestingly inflation continues for about 31 $e$-foldings after the start of the waterfall. The scales associated with the monopoles and cosmic strings are close to each other as shown in Table~\ref{tab:ob1}, with the dimensionless string tension parameter $G\mu = 2.1 \times 10^{-6}$. The cosmic string network is partially but adequately inflated, which makes the gravitational wave spectrum to be in good agreement with the advanced LIGO-VIRGO third run bound, as we show in the next section. We compute $t_e$ by setting the first slow-roll parameter $\epsilon_1\equiv -\dot{H}/ H^2 =1$, where the  slow roll is violated. As an example, we set $T_r= 3 \times 10^{8}$ GeV such that  $t_r= 2.6\times 10^{-24}$ sec. Also, we found that $t_c=1.86\times 10^{-36}$ sec and $t_e= 5.3\times 10^{-36 }$ sec. 

Similarly,  the benchmark point BP2 corresponds to trans-Planckian values of the inflaton $S$, and again inflation continues for 31 $e$-foldings after the start of the waterfall. However, the values of $v_\phi$ and $v_\psi$ are somewhat larger, as shown in Table~\ref{tab:ob1},  yielding $G\mu = 5.4\times 10^{-6}$. For definiteness, we set $T_r= 2 \times 10^{11}$ GeV such that  $t_r= 5.9\times 10^{-30}$ sec, $t_c=1.84\times 10^{-36}$ sec, and $t_e=4.97\times 10^{-36}$ sec. 
\section{Coleman-Weinberg correction to the inflationary potential}\label{sec:RCs}
\begin{table}[h!]
 \centering
 \begin{tabular}{c |  c  }
  \hline \hline
      Fields            &  Squared masses \\
\hline
\hline
    12 gauge bosons   &     $\dfrac{5 g^2}{6} \left( \frac{m_{\phi }^2+  \beta _{S \phi } S^2 }{\beta _{\phi }} \right) $ \\
    \hline\hline 
      12 real scalars   &   $ -m_{\psi }^2+ \beta _{S\psi}S^2 +\frac{\beta _{\psi \phi } \left(m_{\phi }^2+ \beta _{S\phi} \, S^2\right)}{\beta _{\phi }} - 
      \dfrac{\lambda_5 (m_{\phi }^2+  \beta _{S \phi } S^2)}{12 \beta_\phi }$   \\
\hline
 6 real scalars   &   $ -m_{\psi }^2+ \beta _{S\psi}S^2 +\frac{\beta _{\psi \phi } \left(m_{\phi }^2+ \beta _{S\phi} \, S^2\right)}{\beta _{\phi }} - 
      \dfrac{\lambda_5 (m_{\phi }^2+  \beta _{S \phi } S^2)}{6 \beta_\phi }$   \\
\hline
2 real scalars    &   $ -m_{\psi }^2+ \beta _{S\psi}S^2 +\frac{\beta _{\psi \phi } \left(m_{\phi }^2+ \beta _{S\phi} \, S^2\right)}{\beta _{\phi }}$   \\
\hline\hline
8 real scalars    &  $ \frac{\lambda_1 \left(m_\phi^2+\beta_{S\phi} S^2\right)}{24 \beta_\phi } $   \\
\hline
3 real scalars    &  $ \frac{\lambda_1 \left(m_\phi^2+\beta_{S\phi} S^2\right)}{6 \beta_\phi } $   \\
\hline
1 real scalar   &  $ 2 \left(m_{\phi }^2+ \beta _{S\phi}\,S^2 \right)$  \\
\hline
\hline
6 Dirac fermions   &  $ \left( \dfrac{Y_\Phi}{2\sqrt{15 \, \beta_\phi}}\sqrt{ \left(m_{\phi }^2+ \beta _{S\phi}\,S^2 \right)} - 2 Y_S \,S \right)^2 $  \\
\hline
3 Dirac fermions   &  $ \left( \dfrac{2Y_\Phi}{\sqrt{15 \, \beta_\phi}}\sqrt{ \left(m_{\phi }^2+ \beta _{S\phi}\,S^2 \right)} + 2 Y_S \,S   \right)^2 $  \\
\hline
1 Dirac fermion   &  $ \left( \dfrac{3 Y_\Phi}{\sqrt{15\, \beta_\phi}}\sqrt{ \left(m_{\phi }^2+ \beta _{S\phi}\,S^2 \right)} - 2 Y_S \,S   \right)^2 $  \\
  \hline  \hline
  \end{tabular}
 \caption{The  squared masses of scalar and vector fields, as well as the vector-like Dirac fermions during inflation for $S\geq S_c$.}
 \label{tab:minfall}
\end{table}

For completeness we now consider the Coleman-Weinberg (CW) 1-loop correction to the tree level inflation potential in Eq.~(\ref{eq:infpot1}). This correction is given by the standard formula \cite{Coleman:1973jx} 
\begin{equation}
\Delta V_{\rm CW} = \frac{1}{64 \pi^2} \sum_{i} (-1)^{F_i} M_i^4 \ln\left(M_i^2/\Lambda^2 \right) ,
\end{equation}
where $i$ runs over all helicity states, $F_i$ denotes the fermion number and
$M_i^2$ the mass squared of the $i$th state along the inflationary path, and $\Lambda$ is a renormalization scale mass. The mass squared of $S$ and its mixings 
with $\Phi$ give subdominant contributions and can, thus, be ignored. The 
relevant squared masses of gauge boson, $\Phi$, and $\Psi$ components during inflation for $S\geq S_c$ are given in Table~\ref{tab:minfall}. There exist 12 gauge bosons, corresponding to the leptoquarks, with masses squared equal to $\dfrac{5 g^2}{6}  \phi^2 $  ($g$ is the SU(5) gauge coupling constant) originating from the term
\be
g^2 f^{abc} f^{ade} A^b_\mu A^{d \,\mu} \phi^\dagger_c \phi_e ,
\ee  
with $f^{abc}=-2iTr([T^a,T^b]T^c)$ being the SU(5) totally antisymmetric structure constants. The masses squared of the real components of the scalar fields are 
obtained from the potential in Eq.~(\ref{eq:pot_tot}) with the 10-plet providing 20 massive real degrees of freedom and the 24-plet 12 real degrees of freedom.

\begin{figure}[htbp!]
\begin{center}
\subfloat[]
{
    \includegraphics[width=0.75\linewidth]{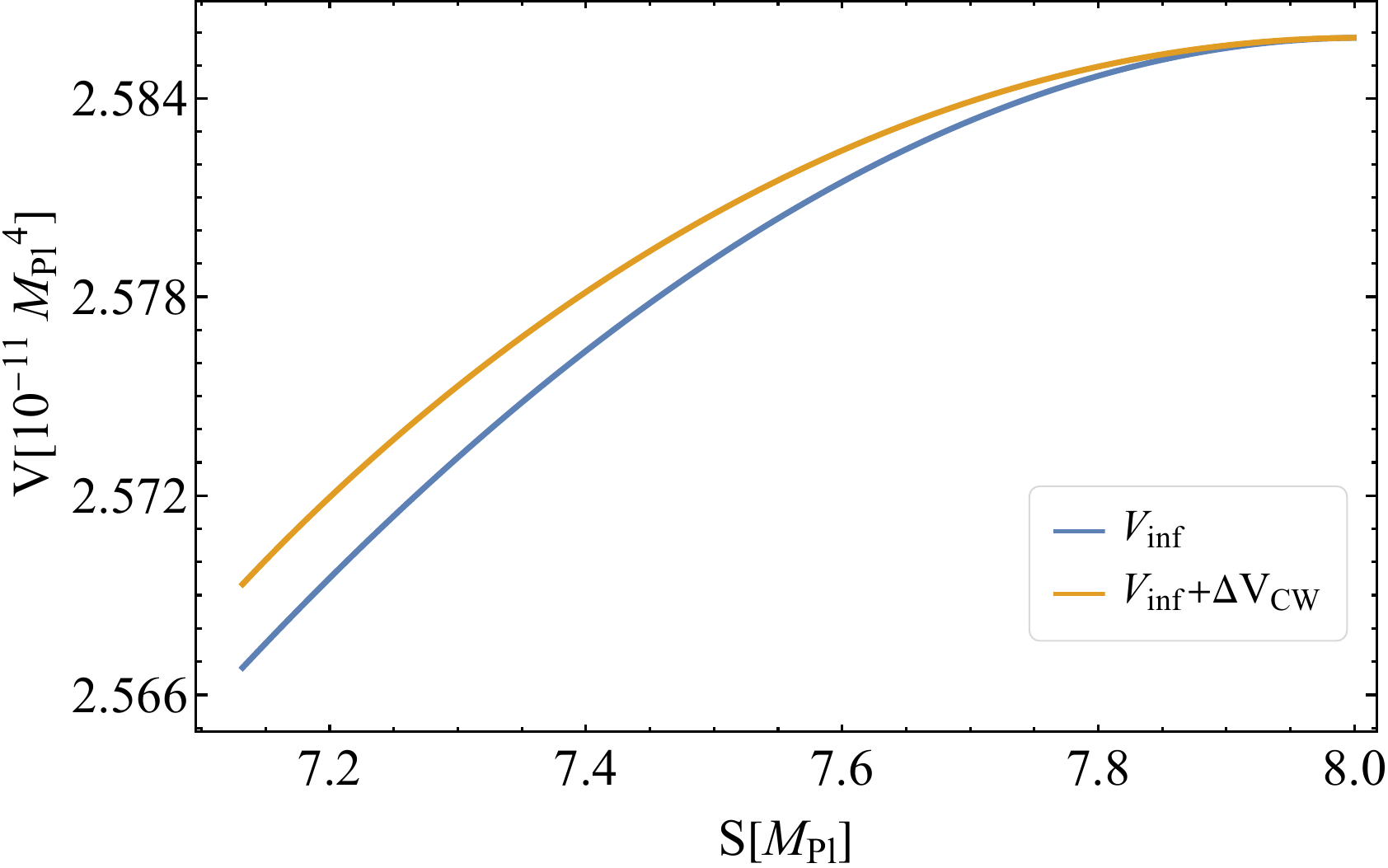}
    \label{fig:Pot-BP1}
}\\
\subfloat[]
{
    \includegraphics[width=0.75\linewidth]{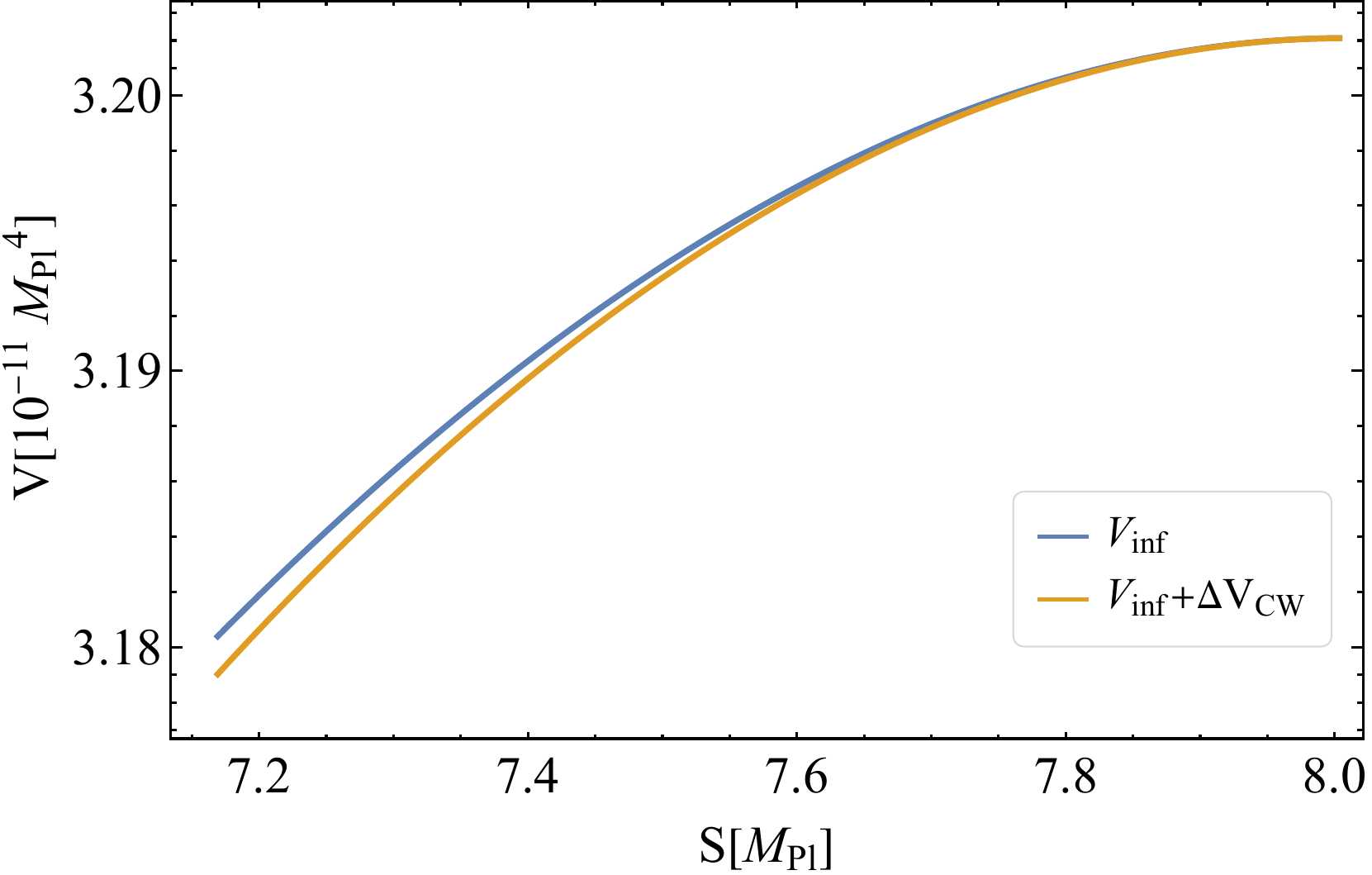}
    \label{fig:Pot-BP2}
}
\end{center}
\caption{The inflation potential for the BP1, where $S_m=8.006 \, M_{\rm Pl}$, $S_*=7.40 \, M_{\rm Pl}$ and $S_c=7.132 \, M_{\rm Pl}$ [panel(a)]. The lower panel corresponds to the BP2, where $S_m=8.006 \, M_{\rm Pl}$, $S_*=7.454 \, M_{\rm Pl}$ and $S_c=7.169 \, M_{\rm Pl}$ [panel(b)]. The blue curve depicts the tree level inflation potential $V_{\rm inf}$ in Eq.~(\ref{eq:infpot1}), while the yellow one represents the total inflation potential including the CW 1-loop correction $\Delta V_{\rm CW}$. 
}\label{fig:Pot}
\end{figure}
%

We chose to impose the renormalization conditions that the CW 
correction and its derivative with respect to $S$ vanish at $S_m$ so that 
the position and height of the hilltop of the tree level potential are 
not affected. To satisfy both these conditions simultaneously, we need 
at least one more adjustable parameter in the CW correction besides $\Lambda$. 
To this end, we introduce an extra pair of fermionic $10_{F_1}$,   
$\overline{10}_{F_2}$ with $X = 4, -4$ respectively so that they do not mix with 
the SM fermions. Their Yukawa couplings $Y_S\, S \, 10_{F1} \overline{10}_{F_2} $ and $Y_\Phi\, \Phi \, 10_{F1} \overline{10}_{F_2} $
provide the extra adjustable parameters ($Y_S,Y_\Phi$), and also give superheavy masses to this vector-like pair. We obtain 10 Dirac fermions with masses squared during inflation also given in Table~\ref{tab:minfall}.

In Fig.~\ref{fig:Pot} we display the tree-level inflation potential in blue, and the total potential including the CW 1-loop correction in yellow, between $S_c$ and $S_m$. For BP1, the solution is $\Lambda= 3.4\times 10^{15} $ GeV, $Y_S=1.18 \times 10^{-4}$ and $Y_\Phi=10^{-4}$, while for BP2 it is  $\Lambda= 6.13\times 10^{15} $ GeV, $Y_S=-1.36 \times 10^{-4}$ and $Y_\Phi=10^{-5}$. We emphasize that the CW radiative correction has a very small contribution for both BPs and slightly improves the inflation observables. 

\section{Cosmic strings and gravitational wave signals at NANOGrav}
\label{sec:string-GW}
{The first step of the symmetry breaking in Eq.~\eqref{eq:SBpatternX} yields monopoles 
that are inflated away,
since $SU(5)$ is broken during inflation. The breaking of $U(1)_Z \times U(1)_X$ to $U(1)_Y$ is achieved by the waterfall 10-plet higgs (with a 
final vev $v_\psi$) and produces cosmic strings.

The procedure of string production is similar to the one described in 
Refs.~\cite{Chakrabortty:2020otp,Lazarides:2021uxv}. As the system crosses $S_c$, $\psi=0$ becomes a maximum, and 
two symmetric minima appear on either side of it. Initially they are 
very shallow and fluctuations from one to the other over $\psi=0$ are very 
frequent. They occur in regions of size the correlation length, which 
is the inverse of the effective mass of $\psi$ at the minimum, and are suppressed in 
accordance with the Ginzburg criterion \cite{GINZBURG} when the energy they 
require exceeds the Hawking temperature of the inflating universe. 
This completes the phase transition and generates the strings since $\psi$   
falls to either of the minima in different regions of size the 
Ginzburg correlation length. Therefore, the initial inter-string 
distance $\xi_{\rm str}$ is expected to be of the same order.

Actually, the transition takes place almost immediately after crossing 
$S_c$ when $|(M_{\rm inf}^2)_{\psi\psi}|$ becomes $\sim H^{2}$, where $H$ is the Hubble 
parameter of the inflating universe. As it turns out the Ginzburg 
correlation length and thus the initial inter-string distance $\xi_{\rm str}$ 
is also of order $H^{-1}$.

After this, $\psi$ moves slowly down its valley of minima and the universe 
continues inflating until $t_e$. Consequently, the string network 
experiences some $e$-foldings ($\Delta N_c$) and the inter-string distance 
becomes super-horizon. During the subsequent field oscillations and 
the radiation dominance that follows, the inter-string distance grows 
proportionally to the scale factor of the universe until $t_F$, where 
the network re-enters the horizon with initial inter-string distance 
of the order of the horizon size, and the strings start generating loops from intercommuting. The system quickly reaches the scaling solution 
which is retained until $t_s$, where the decay of the strings and loops 
by quantum tunneling of monopole-antimonopole pairs starts.

The inter-string separation at time $t_F$ is given by \cite{Chakrabortty:2020otp,Lazarides:2021uxv}
\begin{align}
d_{\rm str}\sim p \xi_{\rm str}\exp\left(\Delta N_c\right)\left(\frac{t_r}{t_e}\right)^{\frac{2}{3}}\left(\frac{t_F}{t_r}\right)^{\frac{1}{2}} ,
\end{align}
where $p \sim 2$ is a geometric factor. The horizon re-entry time of the 
string network is then estimated to be
\begin{align}
t_F\sim \xi_{\rm str}^2 \exp\left(2\Delta N_c\right)t_r^{1/3}t_e^{-4/3} .
\end{align}
}

%
%

  The decay rate per unit length of the string via the production of monopole-antimonpole pairs is given by \cite{Preskill:1992ck,Leblond:2009fq}
\begin{align}\label{eq:Gamma_d}
\Gamma_d = \frac{\mu}{2\pi}\exp(-\pi \kappa),
\end{align}
where $\kappa=m_M^2/\mu$, with $\mu$ being the string tension and $m_M$ the monopole mass. The strings are effectively stable for $\sqrt{\kappa}\gtrsim 9$, and so $\sqrt{\kappa} \approx 8$ is the value usually employed in order to obtain metastable strings.
 In this case, the strings remain effectively stable only before the
cosmic time  $t_s = \frac{1}{\sqrt{\Gamma_d}}$. After $t_s$ they start decaying via monopole-antimonpole pair production. The number density of loops per unit loop length $n(l,t)$ in the radiation dominated Universe is given as \cite{Blanco-Pillado:2013qja,Blanco-Pillado:2017oxo,LIGOScientific:2021nrg}
\begin{align}\label{eq:n-loop-rad}
n_r(l,t) = \frac{0.18 \ \Theta(0.18t-l)}{t^{3/2}(l+\Gamma G\mu t)^{5/2}} ,
\end{align}
where $\Gamma \simeq 50$ is a numerical factor \cite{Vachaspati:1984gt, Vilenkin:2000jqa}. The loop distribution at time $t>t_s$ is given by \cite{Buchmuller:2021mbb,Buchmuller:2023aus} 
\begin{align}\label{eq:n-seg}
n_r(l,t) = \frac{0.18\ \Theta(0.18t-l-\Gamma G\mu(t-t_s))}{t^{3/2}(l+\Gamma G\mu t)^{5/2}}\exp\left[-\Gamma_d\left( l(t-t_s)+\frac{1}{2}G\mu(t-t_s)^2\right)\right] .
\end{align}
We assume that the unresolvable bursts of gravitational waves dominantly from the cusps comprise the stochastic background. The burst rate per unit spacetime volume from the cusps on the loops is expressed as
\begin{align}\label{eq:burst-rate}
\frac{d^2R}{dz \, dl} = N_{\rm cusp} H_0^{-3}\phi_V(z) \frac{2n(l,t(z))}{l(1+z)}\left( \frac{\theta_m(f,l,z)}{2}\right)^2\Theta(1-\theta_m),
\end{align}
 where $N_{\rm cusp}=2.13$ \cite{Cui:2019kkd} is the average number of cusps on a loop and $\theta_m$ is the beam opening angle given by
\begin{align}
\theta_m(f,l,z) = \left[\frac{\sqrt{3}}{4}(1+z)fl\right]^{-1/3}.
\end{align}
%
\begin{figure}[htbp!]
\begin{center}
\subfloat[]
{
    \includegraphics[width=0.75\linewidth]{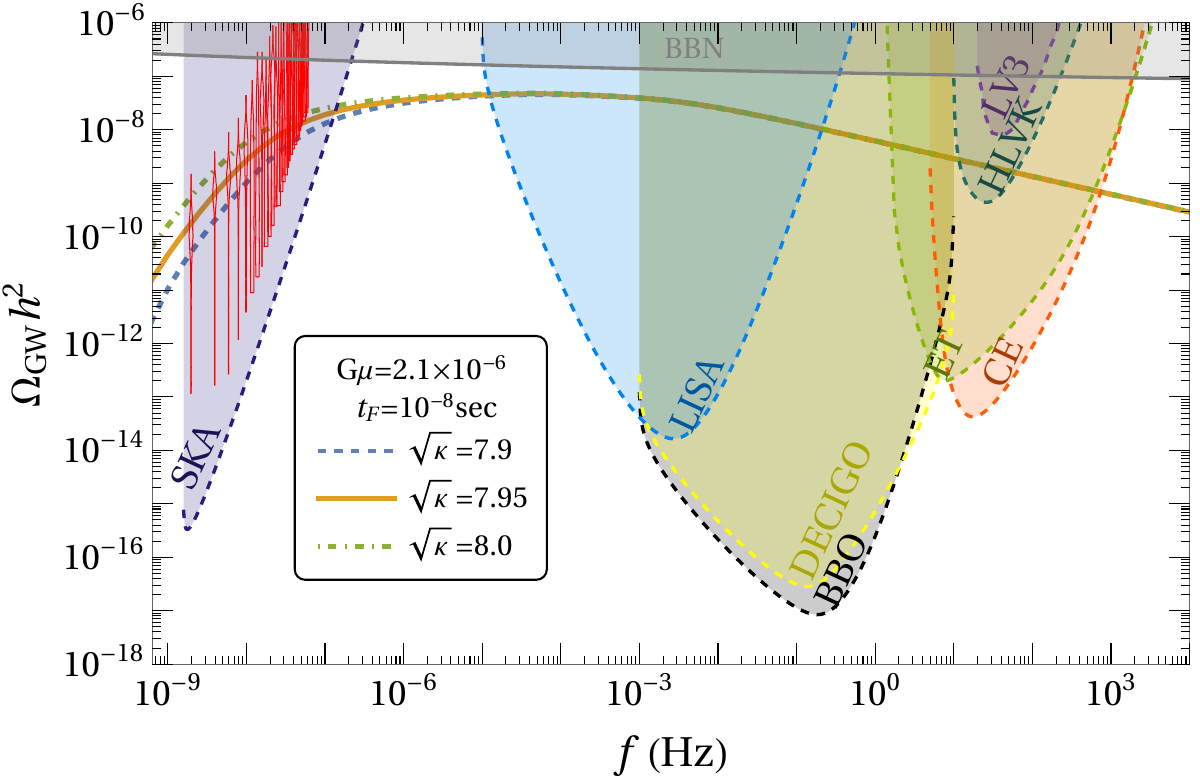}
    \label{fig:gws-1}
}\\
\subfloat[]
{
    \includegraphics[width=0.75\linewidth]{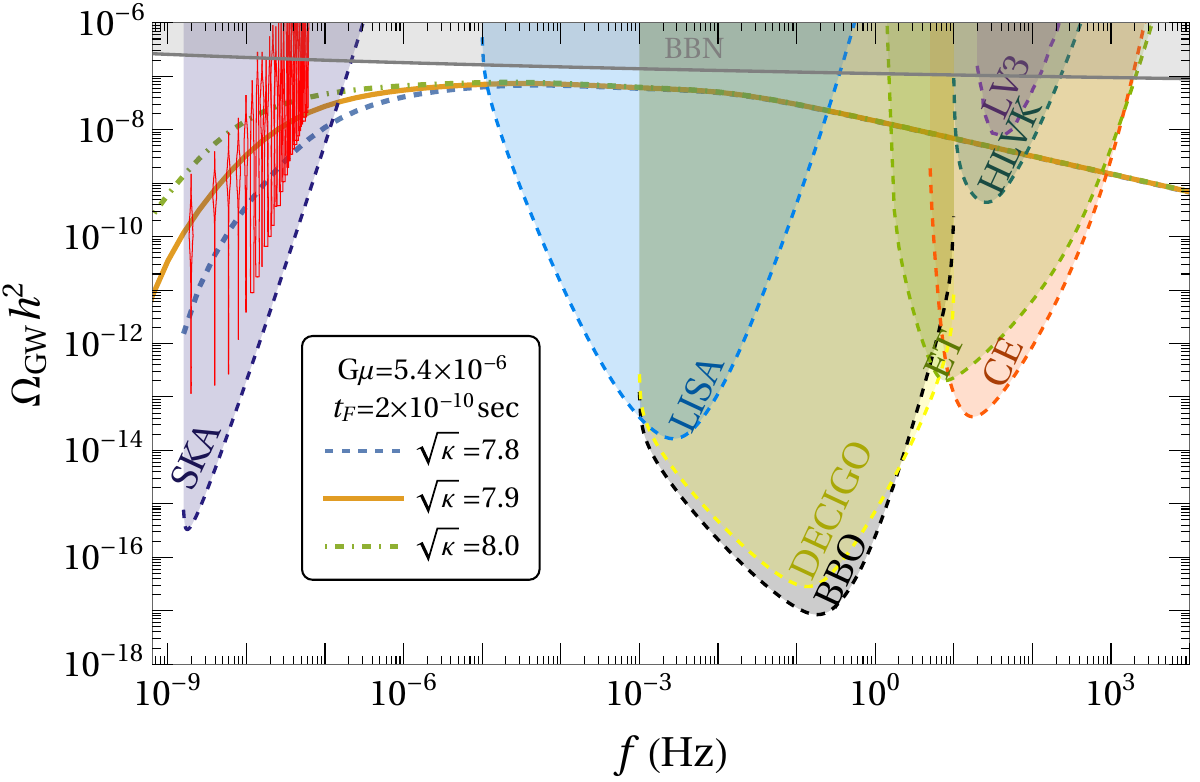}
    \label{fig:gws-2}
}
\end{center}
\caption{Gravitational wave spectrum from metastable cosmic strings with  $G\mu = 2.1\times 10^{-6}$ (Fig.~\ref{fig:gws-1}) and $G\mu = 5.4\times 10^{-6}$ (Fig.~\ref{fig:gws-2}) for varying values of $\sqrt{\kappa}$. The string network experiences about 31 $e$-foldings of inflation, which makes the gravitational wave spectrum consistent with the NANOGrav 15 year data and the third run advanced LIGO-VIRGO (LV3) bound \cite{LIGOScientific:2021nrg}. The posteriors of the HD-correlated free spectra of the NANOGrav data are depicted in the red violin plots as they are presented in Ref.~\cite{NANOGrav:2023hvm}. We also show the power-law integrated sensitivity curves \cite{Thrane:2013oya, Schmitz:2020syl} for several proposed experiments, namely, HLVK \cite{KAGRA:2013rdx}, CE \cite{Regimbau:2016ike}, ET \cite{Mentasti:2020yyd}, DECIGO \cite{Sato_2017}, BBO \cite{Crowder:2005nr, Corbin:2005ny}, LISA \cite{Bartolo:2016ami, amaroseoane2017laser} and SKA \cite{5136190, Janssen:2014dka}, and the Big Bang Nucleosynthesis (BBN) constraint \cite{Mangano:2011ar}.}\label{fig:GWs-MSS}
\end{figure}
%

The time $t$, proper distance $r_p$ and the proper volume element $dV$ can be expressed as functions of cosmological redshift as follows:
\begin{align}\label{eq:t}
t(z)=H_0^{-1}\phi_t(z) \ \ \mathrm{with} \  \ \phi_t(z) = \int_z^\infty\frac{dz'}{\mathcal{H}(z')(1+z')} \ ,
\end{align} 
\begin{align}\label{eq:r}
r_p(z)=H_0^{-1}\phi_r(z) \ \ \mathrm{with} \  \ \phi_r(z) = \int_0^z \frac{dz'}{\mathcal{H}(z')} \ ,
\end{align}
and
\begin{align}
dV(z)=H_0^{-3}\phi_V(z) dz \ \ \mathrm{with} \  \ \phi_V(z) = \frac{4\pi \phi_r^2(z)}{(1+z)^3\mathcal{H}(z)} \ .
\end{align}
Here
\begin{align}\label{eq:mathcalH}
\mathcal{H}(z)=\sqrt{\Omega_{\Lambda,0}+\Omega_{m,0}(1+z)^3+\Omega_{r,0}\mathcal{G}(z)(1+z)^4} \ ,
\end{align}
such that the Hubble parameter in the $\Lambda${CDM} model is given as
\begin{align}
H(z)=H_0\mathcal{H}(z) \ ,
\end{align}
where $H_0$ is the present day value of the Hubble parameter.
In Eq.~\eqref{eq:mathcalH}, $\Omega_{m,0}=0.308$, $\Omega_{r,0}=9.1476\times 10^{-5}$, $\Omega_{\Lambda,0}=1-\Omega_{m,0}-\Omega_{r,0}$ are the fractional energy densities in the Universe from  matter, radiation, and the cosmological constant \cite{Planck:2018vyg}, respectively, $\mathcal{G}(z)$ is given by \cite{Binetruy:2012ze}
\begin{align}
\mathcal{G}(z)=\frac{g_*(z)g_{*S}^{4/3}(0)}{g_*(0)g_{*S}^{4/3}(z)} \, ,
\end{align}
and $g_*$ and $g_{*S}$ are the effective numbers of relativistic degrees of freedom for the energy density and entropy density respectively.

The waveform of the bursts from a cusp is given by \cite{Damour:2001bk}
\begin{align}
h(f,l,z) = g_{1c}\frac{G\mu \, l^{2/3}}{(1+z)^{1/3}r_p(z)}f^{-4/3},
\end{align}
with $g_{1c}\simeq 0.85$ \cite{LIGOScientific:2021nrg}.
The stochastic gravitational wave background at frequency $f$ is expressed as \cite{Olmez:2010bi, Auclair:2019wcv, Cui:2019kkd, LIGOScientific:2021nrg}{
\begin{align}\label{eq:GWs-Omega-cusps}
\Omega_{GW}(f) = \frac{4\pi^2}{3H_0^2}f^3\int_{z_*}^{z(t_F)} dz \int dl \, h^2(f,l,z)\frac{d^2R}{dz \, dl} \ ,
\end{align}
where the lower limit on $z$ is obtained from
\begin{align}
f = \int_{0}^{z_*} dz \int dl \frac{d^2R}{dz \, dl}
\end{align}
and eliminates the contribution of infrequent bursts from the stochastic gravitational wave background. The integration limits on $l$ are generally taken from zero to the size of the particle horizon, which is equal to $2t$ ($3t$) for radiation (matter) domination. However, 
the theta functions in Eqs.~\eqref{eq:n-loop-rad}, \eqref{eq:n-seg} and \eqref{eq:burst-rate} select the proper limits.}
There are several papers in the literature that have previously discussed the gravitational waves from hybrid and metastable cosmic strings including Refs.~\cite{Martin:1996ea,Martin:1996cp,Leblond:2009fq,
Buchmuller:2019gfy,Buchmuller:2020lbh,
Buchmuller:2021dtt,Buchmuller:2021mbb,Masoud:2021prr,Dunsky:2021tih,Ahmed:2022rwy,Afzal:2022vjx,Lazarides:2022jgr}.

{In Fig.~\ref{fig:GWs-MSS} we show the gravitational wave spectrum for $G\mu = 2.1\times 10^{-6}$ (Fig.~\ref{fig:gws-1}) and $G\mu = 5.4\times 10^{-6}$ (Fig.~\ref{fig:gws-2}) corresponding to the BP1 and BP2 for different values of $\sqrt{\kappa}$. In both cases, the string network experiences around 31 $e$-foldings of inflation which enables us to satisfy the advanced LIGO-VIRGO third run bound.}   
Indeed, thanks to inflation, $G\mu \sim  10^{-6}$ is consistent with the NANOGrav 15 year data and the third advanced LIGO-VIRGO bound \cite{LIGOScientific:2021nrg}.\footnote{{ The gravitational wave spectra exhibit an $f^2$ power law at low frequencies, followed by an almost scale-invariant plateau, and an $f^{-1/3}$ tail in the high-frequency region depending on $t_F$, and agree with Refs.~\cite{Buchmuller:2021mbb,Buchmuller:2023aus} that employ a self-consistent method for computing these spectra. 
This method is compatible with other self-consistent methods used in 
the literature.}}

Before concluding this section, a few remarks about the parameter $\kappa$ are in order. With $\sqrt{\kappa}= m_M/\sqrt{\mu}$ and setting $m_M \sim \dfrac{4\pi}{g}v_\phi$,  $\sqrt{\mu } \sim \sqrt{\pi} v_\psi$, where $g$ denotes the $SU(5)$ gauge coupling, we find that $\alpha_G \equiv g^2/4\pi \sim 0.2$ for both BPs.
%
\section{Proton lifetime}
\label{sec:proton_life_time}
The exchange of superheavy gauge bosons $(X',Y')$ with the SM quantum numbers $(3,2,-1/6)$ allows the  proton to decay into $e^+ \, \pi^0$ in flipped $SU(5)$. After integrating out $(X',Y')$, the relevant operator in the physical basis is expressed as \cite{FileviezPerez:2004hn,Dorsner:2004xx,Nath:2006ut}
\begin{align}\label{operator_physical_basis}
\mathcal{O}_R^{d=6}\left( e, d^c\right) & = \mathcal{W_C} \epsilon^{ijk} \overline{u^c_i}\gamma^\mu u_j \overline{d^c_{k}} \gamma_\mu e \ , 
\end{align}
with the Wilson coefficient given by
\begin{align}
\mathcal{W_C} & \approx \frac{|V_{ud}|^2}{v_\phi^2} \ , 
\end{align}
where $|V_{ud}| = 0.9742$ is the Cabibbo–Kobayashi–Maskawa matrix element \cite{Tanabashi:2018oca}. The partial lifetime for the decay channel $p\to \pi^0 e^+$ is given by \cite{Chakrabortty:2019fov}
\begin{align}
\tau_p = & \Bigg[ \frac{m_p}{32\pi}\left(1-\frac{m_{\pi^0}^2}{m_p^2}\right)^2 A_L^2
 A_S^2\mathcal{W_C}^2 |\langle \pi^0 \rvert (ud)_R u_L\lvert p \rangle |^2  \Bigg]^{-1} ,
\end{align}
where $m_p$ and $m_{\pi^0}$ denote the proton and pion masses respectively. $A_L\approx 1.25$ is the long-range (from the electroweak scale to the QCD scale) \cite{Nihei:1994tx}, and $A_S$ is the short-range (from GUT to the electroweak scale) enhancement factor \cite{Buras:1977yy}. We have taken $A_L=2.0$ and the matrix element $\langle \pi^0 \rvert (ud)_R u_L\lvert p \rangle = -0.131$ from lattice computation \cite{Aoki:2017puj}. 

The partial lifetime of the proton for  benchmark points BP1 and BP2 in Table~\ref{tab:ob1} are estimated to  be {$\tau_p\approx 2.1 \times 10^{36}$ and $1.3\times 10^{37}$ yrs, repectively, which lie above the bound from Super-Kamiokande \cite{Super-Kamiokande:2020wjk} and beyond the reach of the Hyper-Kamiokande \cite{Dealtry:2019ldr} experiment.}
%
\section{Conclusions}
\label{sec:conc}
We have explored how in a non-supersymmetric hybrid inflation model based on flipped $SU(5)$ superheavy metastable strings appear after inflation and produce a stochastic gravitational wave spectrum which is compatible with the NANOGrav 15 year data. The symmetry breaking $SU(5) \times U(1)_X$ to $SU(3)_c \times SU(2) \times U(1)_Z \times U(1)_X$ {at a scale of order $10^{16}$ GeV} produce monopoles which are inflated away. The superheavy metastable strings are associated with the symmetry breaking of $U(1)_Z \times U(1)_X$, with the dimensionless string tension parameter {$G \mu\sim   10^{-6}$.} The string network experiences an adequate number of $e$-foldings of inflation such that the gravitational wave background is compatible with the third advanced LIGO-VIRGO bound.
The scalar spectral index {$n_s \simeq 0.963$}, and the tensor-to-scalar ratio is predicted to lie in the range $r \sim  10^{-4}- 10^{-3}$. Our flipped $SU(5)$ model of metastable strings predicts the proton lifetime to lie in the range {$\tau_p\approx  10^{36}-10^{37}$ yrs.}
\acknowledgments
This work is supported by the Hellenic Foundation for Research and Innovation (H.F.R.I.) under the “First Call for H.F.R.I. Research Projects to support Faculty Members and Researchers and the procurement of high-cost research equipment grant” (Project Number: 2251) (G.L. and Q.S.) and the National Research Foundation of Korea grant by the Korea government: 2022R1A4A5030362 (R.M.).
\bibliographystyle{mystyle}
\bibliography{GUT_TD.bib}

\end{document}